\documentclass[11pt]{article}

\usepackage[margin=1in]{geometry}
\usepackage{times}
\usepackage{latexsym}
\usepackage{amsmath}
\usepackage{amssymb}
\usepackage{booktabs}
\usepackage{multirow}
\usepackage{graphicx}
\usepackage{float}
\usepackage{subcaption}
\usepackage{xcolor}
\usepackage{url}
\usepackage[hidelinks]{hyperref}
\usepackage[capitalize,noabbrev]{cleveref}
\usepackage[numbers,sort&compress]{natbib}

\newcommand{\R}{\mathbb{R}}

\newcommand{\x}{\mathbf{x}}
\newcommand{\z}{\mathbf{z}}

\newcommand{\y}{\mathbf{y}}

\newcommand{\cvec}{\mathbf{c}}
\newcommand{\smiles}{\mathrm{SMILES}}
\newcommand{\graph}{\mathcal{G}}

\newcommand{\drugs}{\mathcal{D}}
\newcommand{\loss}{\mathcal{L}}
\newcommand{\lossexpr}{\mathcal{L}_{\mathrm{expr}}}
\newcommand{\losscc}{\mathcal{L}_{\mathrm{cc}}}
\newcommand{\lossadv}{\mathcal{L}_{\mathrm{adv}}}

\newcommand{\pos}[1]{\textcolor{green!45!black}{#1}}

\DeclareMathOperator{\concat}{concat}

\title{scCycleMol: Modeling Cell-Cycle-Aware Single-Cell Drug Perturbation Responses}

\author{
Dingping Zhao\textsuperscript{*}\\
\small Division of Pharmacognosy, School of Pharmaceutical Sciences,\\
\small State Key Laboratory of Natural and Biomimetic Drugs, Peking University, Beijing, China
\and
Jie Lin\textsuperscript{*}\\
\small Department of Computer Science at School of Informatics,\\
\small Xiamen University, Xiamen, China
\and
Feng Xu\textsuperscript{\dag}\\
\small Division of Pharmacognosy, School of Pharmaceutical Sciences,\\
\small State Key Laboratory of Natural and Biomimetic Drugs, Peking University, Beijing, China
\and
Zhengwei Xie\textsuperscript{\dag}\\
\small Peking University Health Science Center, Peking University, Beijing, China\\[0.5em]
\small \textsuperscript{*}Equal contribution. \textsuperscript{\dag}Corresponding authors: \href{mailto:xufeng76@hsc.pku.edu.cn}{xufeng76@hsc.pku.edu.cn}, \href{mailto:xiezhengwei@hsc.pku.edu.cn}{xiezhengwei@hsc.pku.edu.cn}
}
\date{}

\begin{document}
\maketitle

\begin{abstract}
Single-cell drug perturbation models should capture transcriptional response magnitude and whether a treatment changes the proliferative state of the cell.
This is difficult because cell-cycle variation is often treated as a nuisance factor, and benchmark processing rarely makes drug-induced phase changes a first-class prediction target.
We introduce scCycleMol, a cell-cycle-aware perturbation prediction framework built on a curated 24-hour SciPlex3 benchmark with standardized molecule identities, dose and cell-line metadata, modeled genes, and expression-derived cell-cycle supervision.
scCycleMol derives cell-cycle supervision from the treated state and applies it to the predicted treated expression without using phase as an input covariate.
The main model uses a learnable full-expression cell-cycle head with circular G1/S/G2M phase targets, and we evaluate both readout-only supervision with stop-gradient and closed-loop supervision that backpropagates through the decoder, dose-response module, and molecular drug representation.
We also evaluate molecular representation choices and pretraining sources to separate cell-cycle objective effects from representation and preprocessing effects.
On a processed 24-hour SciPlex3 benchmark with 635,541 cells, 186 non-control molecule-level perturbation units, 188 valid label-level compound embeddings, three cancer cell lines, four nonzero treatment doses plus DMSO control, and 5,080 modeled genes, the best LINCS-pretrained circular variant reaches 0.9093 mean all-gene $r^2$ and 0.6843 mean DE-gene $r^2$.
Under matched preprocessing, closed-loop cell-cycle supervision raises phase accuracy by 0.54--0.62 absolute points while keeping mean all-gene $r^2$ within 0.003 of matched chemCPA no-cell-cycle models; Tahoe-pretrained readout-only circular supervision gives the strongest supervised phase readout, reaching 0.9609 phase accuracy.

\end{abstract}

\section{Introduction}

Drug perturbations do not merely shift average gene expression.
They can arrest cells, accelerate transitions, or redistribute a population across cell-cycle phases.
Live-cell studies of anti-cancer agents show drug- and dose-specific changes in cell-cycle phasing~\citep{gross2023phase}, while long-term single-cell imaging under cisplatin shows that proliferation status can shape arrest-versus-death outcomes~\citep{granada2020proliferation}.
For single-cell perturbation prediction, this matters: a model can reconstruct a plausible mean transcriptome while missing whether a drug pushes cells toward S phase, G2/M, or another proliferative state.
Such errors directly affect how we interpret drug mechanism, response heterogeneity, and out-of-distribution response.

Recent single-cell perturbation models, including conditional autoencoder and compositional perturbation frameworks, have made substantial progress in predicting transcriptional responses from cell state, drug identity, dose, and covariates~\citep{lotfollahi2019scgen,lotfollahi2023cpa,hetzel2022chemcpa}.
At the same time, recent benchmarks caution that perturbation-effect prediction remains difficult: gene-perturbation models do not always outperform simple linear baselines, and broad single-cell benchmarks emphasize generalization across unseen contexts and perturbation settings~\citep{ahlmann2025linearbaselines,wei2026benchmarking}.
Two limitations are especially important for drug response modeling.
First, cell-cycle variation is often left implicit, treated as an ordinary covariate, or removed as unwanted variation, even when drug-induced phase changes are part of the biological response to be predicted.
Second, benchmark preprocessing often emphasizes expression reconstruction while leaving cell-cycle labels, molecule identities, and pretraining resources weakly connected.
This makes it difficult to ask whether a perturbation model predicts drug-induced proliferation state rather than only a plausible transcriptome.

We propose scCycleMol, a cell-cycle-aware framework for single-cell drug perturbation prediction.
The model builds on a ComPert/chemCPA-style conditional autoencoder, but changes the learning signal around the predicted treated state.
Instead of feeding cell-cycle phase to the model as an input covariate, scCycleMol derives supervision from treated single-cell profiles and applies it to the predicted treated expression.
This design asks the model to explain drug-induced proliferation changes through its predicted response rather than by conditioning on the observed phase label.
To make this supervision dense and differentiable, scCycleMol attaches a lightweight cell-cycle head to the full predicted expression vector and trains it with circular G1/S/G2M phase targets.
We evaluate this head as a readout-only module with stop-gradient and as a closed-loop module whose loss updates the upstream decoder, dose-response module, and molecular drug encoder.
In the closed-loop setting, the molecular representation predicts a perturbation response, and the predicted biological state gives feedback on whether that response is cell-cycle consistent.

scCycleMol also studies how drug representation and pretraining interact with this biological supervision.
The drug representation can use SMILES language embeddings, molecular graph embeddings, or their projected combination, but we treat these choices as empirical components within the cell-cycle-aware prediction task.
Our embedding ablations show that the proposed embedding improves LINCS-pretrained OOD expression metrics over the chemCPA embedding, while SciPlex-pretrained embeddings are nearly tied and differ mainly in variance preservation.
This motivates joint reporting of expression response, DE-gene response, variance preservation, and phase accuracy.

Our contributions are:
\begin{enumerate}
    \item We organize a cell-cycle-aware drug perturbation benchmark from 24-hour SciPlex3, with standardized molecule identities, dose and cell-line metadata, modeled genes, phase labels, and compatible LINCS and Tahoe pretraining resources.
    \item We formulate treated-state cell-cycle supervision for single-cell drug perturbation prediction, using a learnable full-expression head and circular G1/S/G2M phase targets with readout-only and closed-loop variants while withholding phase from the input covariates.
    \item We separate preprocessing and objective effects by reporting raw CPA/chemCPA reference rows, matched chemCPA no-cell-cycle baselines under our processed contract, and cell-cycle-aware scCycleMol variants. Under matched preprocessing, closed-loop supervision raises phase accuracy by 0.54--0.62 absolute points, while the readout-only circular variant gives a stronger supervised phase readout.
\end{enumerate}

\section{Related Work}
\label{sec:related}

\subsection{Single-Cell Perturbation Prediction}

Single-cell perturbation models seek to predict the transcriptional response of a cell under genetic, chemical, or environmental interventions~\citep{zhang2026deep,zhang2026cistranscell,jiang2026learning}.
Latent translation methods such as scGen model perturbation effects as transformations in a learned expression space~\citep{lotfollahi2019scgen}.
Compositional perturbation models and CPA-style approaches extend this idea by separating basal cell state, perturbation identity, dose, and covariates, often using adversarial objectives to encourage disentanglement~\citep{lotfollahi2023cpa}.
chemCPA further incorporates chemical representations to support drug generalization~\citep{hetzel2022chemcpa}.
Related approaches address complementary regimes.
GEARS models combinatorial genetic perturbations with graph priors~\citep{roohani2024gears}, CellOT learns perturbation responses as neural optimal transport between single-cell distributions~\citep{bunne2023cellot,chen2025fast}, and PRnet and PerturbNet target chemical or mixed chemical/genetic generalization to unseen perturbations~\citep{qi2024prnet,yu2025perturbnet,liu2025learning}.
Recent diffusion-based models further frame single-cell drug-response and perturbation prediction as conditional generative modeling of treated cell states~\citep{liang2025scppdm,shi2026statexdiff}.
scCycleMol builds on this family of conditional response predictors, but differs in the target of supervision: instead of optimizing only expression reconstruction and disentanglement, it requires predicted treated expression to preserve a specific biological state axis, namely cell cycle.
Chemical representation remains important for drug generalization~\citep{pegoraro2026role}.
SMILES language models, molecular graph neural networks, and multi-view molecular learning capture complementary chemical information~\citep{chithrananda2020chemberta,xu2019gin,rong2020grover,zhang2024mvmrl}.
In this work, these representations are evaluated as components of a cell-cycle-aware perturbation model rather than as the central object of study.

\subsection{Cell-Cycle State Modeling and Perturbation Resources}

Cell cycle is a central source of variation in single-cell transcriptomics.
Standard workflows often score cells using curated S and G2/M cell-cycle gene sets from prior single-cell studies and assign coarse G1/S/G2M labels, with implementations in Seurat and Scanpy~\citep{tirosh2016melanoma,satija2015seurat,wolf2018scanpy}.
Depending on the scientific question, cell cycle may be corrected as a confounder or retained as a biological signal.
In drug perturbation response, the latter is often crucial: many compounds alter proliferation, arrest, and phase distributions.
scCycleMol therefore reframes cell cycle as a prediction target.
This differs from using phase labels as covariates, because the model cannot explain away drug-induced state changes by conditioning on the answer.

Large-scale resources such as LINCS L1000, SciPlex, and more recent high-throughput single-cell drug screens provide complementary regimes of scale, modality, and biological resolution~\citep{subramanian2017lincs,srivatsan2020sciplex,zhang2025tahoe100m}.
Bulk or reduced-gene resources offer broad chemical coverage, while single-cell resources expose heterogeneity and state redistribution.
For cell-cycle-aware perturbation modeling, these resources must be connected through consistent molecule identities, gene vocabularies, treatment metadata, and phase annotations.
Accordingly, scCycleMol uses processed SciPlex3 as the primary OOD benchmark and uses L1000 and Tahoe-100M as broader resources for pretraining and cross-resource analysis where compatibility supports direct comparison.

\section{Method}
\label{sec:method}

\begin{figure}[t]
    \centering
    \includegraphics[width=\linewidth]{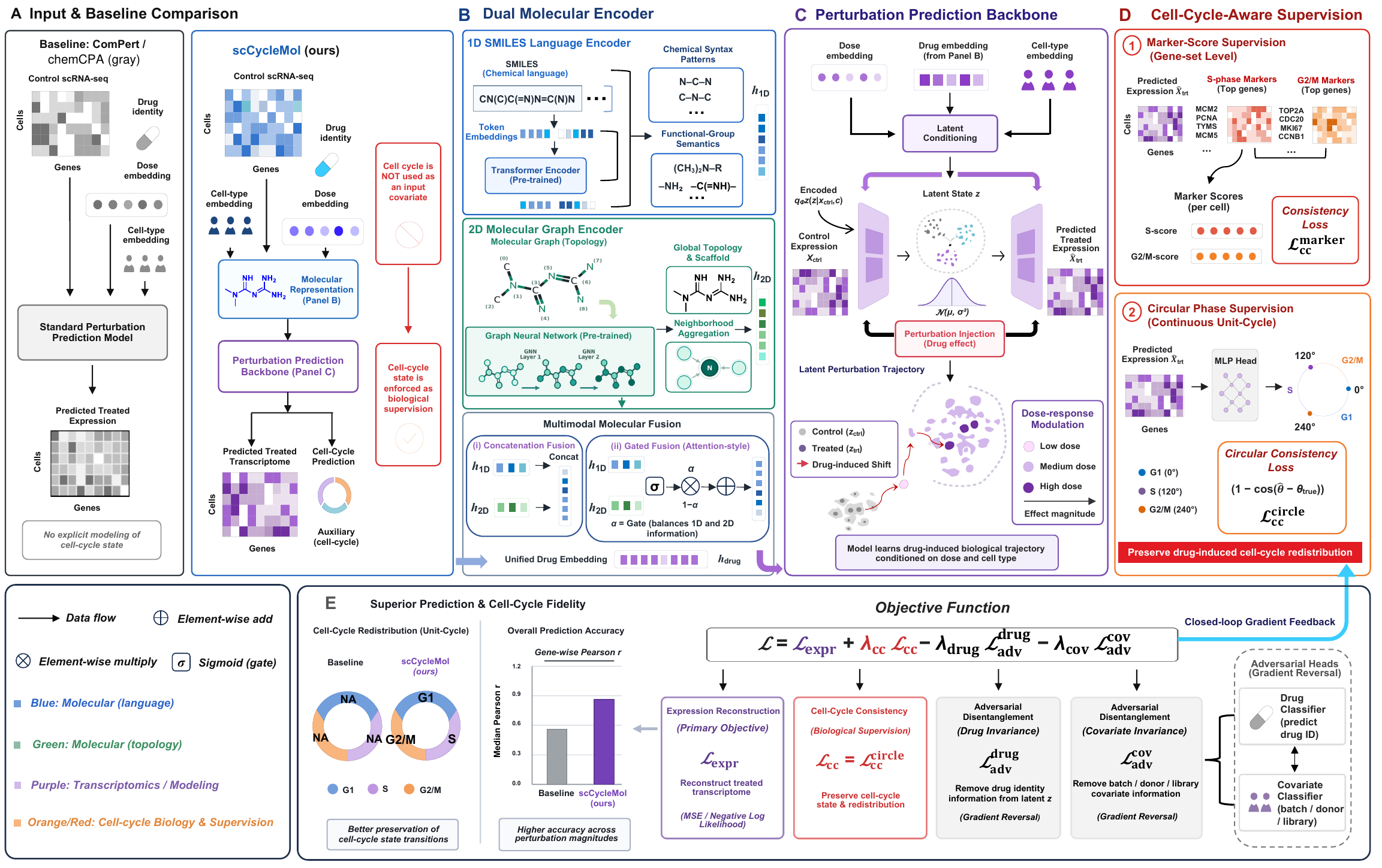}
    \caption{Overview of scCycleMol. The model encodes basal single-cell expression, conditions on drug and dose, and applies closed-loop cell-cycle supervision to the predicted treated state.}
    \label{fig:method_overview}
\end{figure}

\subsection{Problem Setup}

Let $\x_i \in \R^G$ denote the basal or control gene-expression profile of cell $i$ over $G$ genes.
Each training example is associated with a drug $d_i \in \drugs$, dose $u_i$, and biological covariates such as cell type $\cvec_i$.
The target is treated expression $\y_i \in \R^G$.
Unlike standard perturbation prediction, we also observe or derive cell-cycle supervision from the treated cell, denoted $s_i$.
In the current model, $s_i$ is a phase label in $\{\mathrm{G1}, \mathrm{S}, \mathrm{G2M}\}$ represented through a circular encoding.
The goal is to learn a predictor
\begin{equation}
    \hat{\y}_i = f_\theta(\x_i, d_i, u_i, \cvec_i)
\end{equation}
that accurately predicts treated expression while preserving the cell-cycle state of the treated population.
Cell-cycle information is used only as supervision on the predicted treated state, not as an input covariate.
This prevents the model from conditioning on the observed phase label and instead requires drug-induced proliferation or phase changes to be represented through the predicted expression profile.

\subsection{Base Perturbation Model}

scCycleMol follows the conditional autoencoder design used in ComPert/chemCPA-style models~\citep{lotfollahi2023cpa,hetzel2022chemcpa}.
An encoder maps control expression into a latent basal state,
\begin{equation}
    \z_i = E_\theta(\x_i).
\end{equation}
A drug encoder maps drug identity and molecular representation into a perturbation vector, optionally modulated by a dose-response function.
Cell-type or other covariate embeddings are added in the latent space.
The decoder then produces predicted treated expression:
\begin{equation}
    \hat{\y}_i = D_\theta(\z_i + r_\theta(d_i, u_i) + q_\theta(\cvec_i)).
\end{equation}
The expression loss is
\begin{equation}
    \lossexpr = \frac{1}{N}\sum_{i=1}^{N} \ell_{\mathrm{expr}}(\hat{\y}_i, \y_i),
\end{equation}
where $\ell_{\mathrm{expr}}$ can be mean squared error or the reconstruction objective used by the base model.
When inherited from ComPert/chemCPA, adversarial objectives are used to discourage unwanted leakage of drug or covariate information into the basal latent representation.

\subsection{Cell-Cycle-Aware Treated-State Supervision}

Earlier score-extraction variants computed cell-cycle supervision from a small fixed subset of S and G2/M genes.
That route made the auxiliary gradient sparse and tied the loss to a hand-defined scoring rule.
To provide a denser and more flexible supervisory signal, scCycleMol uses a learnable cell-cycle head that operates on the full predicted treated expression profile.

\paragraph{Cell-cycle prediction head.}
The main model replaces fixed score extraction with a lightweight learnable head.
Given the predicted treated expression $\hat{\y}_i \in \R^G$, the head
\begin{equation}
    \hat{p}_i = H_\theta(\hat{\y}_i)
\end{equation}
maps the full expression vector to a two-dimensional phase representation.
We implement $H_\theta$ as a small multilayer perceptron with a 64-dimensional hidden layer.
Because the head receives all predicted genes rather than a fixed gene subset, it provides a dense phase readout of the predicted treated profile while adding only a small parameter overhead.

\paragraph{Circular phase encoding.}
Discrete phase labels impose an artificial ordering if treated as class indices.
We therefore map the three coarse phases to equally spaced prototypes on the unit circle:
\begin{equation}
    \mathrm{G1}\mapsto (1,0),\quad
    \mathrm{S}\mapsto \left(\cos \frac{2\pi}{3}, \sin \frac{2\pi}{3}\right),\quad
    \mathrm{G2M}\mapsto \left(\cos \frac{4\pi}{3}, \sin \frac{4\pi}{3}\right).
\end{equation}
The circular loss is
\begin{equation}
    \losscc^{\mathrm{circ}} =
    \frac{1}{N}\sum_{i=1}^{N}
    \left\|\hat{p}_i - p_i\right\|_2^2,
\end{equation}
where $p_i$ is the unit-circle target.
At evaluation time, the predicted phase is obtained by nearest prototype on the circle.

\paragraph{Readout-only and closed-loop supervision.}
The readout-only circular variant attaches the cell-cycle head to a stop-gradient copy of the predicted treated expression, $H_\theta(\mathrm{sg}(\hat{\y}_i))$.
In this setting, the circular phase loss trains the phase readout but does not update the decoder, dose-response module, or drug-conditioned perturbation pathway.
The closed-loop circular variant removes this stop-gradient operation.
Therefore, the circular phase loss backpropagates through the decoder, dose-response module, and drug-conditioned perturbation pathway.
This creates a closed-loop training signal: the drug-conditioned perturbation pathway determines the predicted response, and the predicted biological state provides feedback on whether that response is cell-cycle consistent:
\begin{equation}
    (d_i,u_i) \rightarrow h_d \rightarrow r_\theta(d_i,u_i)
    \rightarrow \hat{\y}_i \rightarrow H_\theta(\hat{\y}_i)
    \rightarrow \hat{p}_i .
\end{equation}

\subsection{Drug Representation and Pretraining Variants}

The drug encoder is trained in the context of treated-state prediction rather than standalone molecular property prediction.
Its representation must support both expression reconstruction and cell-cycle-consistent state prediction.
We therefore evaluate molecular views before passing the drug embedding into the perturbation predictor.
The 1D branch encodes the SMILES string with a chemical language model:
\begin{equation}
    z^{1d}_d = \mathrm{LM}_\phi(\smiles_d).
\end{equation}
This representation captures token context, substructure syntax, and chemical language regularities.
The 2D branch constructs a molecular graph $\graph_d$ from the same molecule and encodes it with a graph neural network:
\begin{equation}
    z^{2d}_d = \mathrm{GNN}_\psi(\graph_d),
\end{equation}
capturing atom-bond neighborhoods and topology.
Both representations are projected into the perturbation model's drug space:
\begin{equation}
    h^{1d}_d = W_{1d} z^{1d}_d,\qquad
    h^{2d}_d = W_{2d} z^{2d}_d.
\end{equation}

We consider two fusion strategies.
Static concatenation forms
\begin{equation}
    h^{\mathrm{cat}}_d = W_f \concat(h^{1d}_d, h^{2d}_d).
\end{equation}
Gated fusion learns a molecule-specific gate,
\begin{equation}
    g_d = \sigma\left(\mathrm{MLP}_g(\concat(h^{1d}_d, h^{2d}_d))\right),
\end{equation}
and combines views as
\begin{equation}
    h^{\mathrm{gate}}_d = g_d \odot h^{1d}_d + (1-g_d)\odot h^{2d}_d.
\end{equation}
Let $h_d$ denote the selected drug representation, which may be a single projected view or a combined representation.
This drug embedding is passed to the dose-response module and downstream perturbation predictor.
Because the downstream losses are applied to both predicted expression and predicted cell-cycle phase, the gate is optimized to select molecular views that are useful for biological perturbation response rather than molecule-only objectives.

\subsection{Training Objective}

The full objective is
\begin{equation}
    \loss =
    \lossexpr
    + \lambda_{\mathrm{cc}}\losscc
    - \lambda_{\mathrm{drug}}\lossadv^{\mathrm{drug}}
    - \lambda_{\mathrm{cov}}\lossadv^{\mathrm{cov}},
\end{equation}
Here, $\lossadv^{\mathrm{drug}}$ is the loss of a drug adversary that predicts perturbation identity from the basal latent state $\z_i$ using multi-label binary cross-entropy.
The term $\lossadv^{\mathrm{cov}}$ sums covariate-adversary losses that predict nuisance covariates such as cell type from $\z_i$ using cross-entropy.
The adversary networks are optimized to minimize these prediction losses, whereas the encoder-side objective includes them with negative signs, making $\z_i$ less informative about drug and covariate labels.
In the circular variants, $\losscc=\losscc^{\mathrm{circ}}$.
The readout-only circular variant applies this loss after stop-gradient and isolates the value of the full-expression circular phase head as a supervised readout.
The closed-loop circular variant allows gradients from the same loss to update the upstream perturbation pathway, isolating the value of biological-state feedback.
Thus, scCycleMol couples drug-conditioned perturbation prediction with treated-state cell-cycle supervision: the drug embedding predicts the expression shift, while the circular phase head either reads out or feeds back whether that shift induces the correct cell-cycle state.

\section{Experiments}
\label{sec:experiments}

Our experiments establish three findings.
First, we separate raw CPA/chemCPA reference rows from models evaluated under our processed molecule-aware SciPlex3 contract.
Second, closed-loop cell-cycle supervision adds a targeted phase-consistency signal: it raises phase accuracy by 0.54--0.62 absolute points relative to matched chemCPA no-cell-cycle models while keeping expression reconstruction nearly unchanged.
Third, readout-only circular supervision and pretraining source control complementary parts of the trade-off: LINCS pretraining favors expression metrics, whereas Tahoe pretraining gives the highest supervised phase readout.

\subsection{Datasets and Splits}

\paragraph{SciPlex3.}
SciPlex3 is the primary single-cell drug perturbation benchmark~\citep{srivatsan2020sciplex}.
We use a processed 24-hour benchmark with 635,541 cells, 186 non-control molecule-level perturbation units, 188 valid label-level compound embeddings, three cancer cell lines, four nonzero treatment doses plus DMSO control, and 5,080 modeled genes to evaluate expression prediction, DE-gene response prediction, phase accuracy, and out-of-distribution generalization.
The split and evaluation units are molecule-level identities, whereas label-level compound embeddings are retained as molecular representation inputs and traceability entries.
The preprocessing contract defines model-facing perturbation identities at the molecule level from RDKit-standardized SMILES~\citep{rdkit}; original product labels are retained as traceability metadata rather than used as drug identities.
Following the chemCPA SciPlex3 convention~\citep{hetzel2022chemcpa}, we use a chemCPA-compatible OOD fine-tuning split.
The train and test partitions support internal validation and model selection, whereas the OOD partition is reserved for formal unseen-molecule evaluation.

\paragraph{LINCS pretraining.}
LINCS L1000 provides broader chemical coverage and is used as a pretraining source for compatible perturbation representations~\citep{subramanian2017lincs}.
We evaluate whether LINCS initialization improves the downstream SciPlex3 expression and cell-cycle metrics after fine-tuning.
Before formal SciPlex3 OOD evaluation, the L1000 pretraining input is filtered to remove molecules held out in the SciPlex3 OOD partition, while source perturbation identifiers are retained only for traceability.

\paragraph{Tahoe pretraining.}
Tahoe-100M is used as a large-scale single-cell pretraining source~\citep{zhang2025tahoe100m}.
It provides a complementary pretraining regime with substantially larger single-cell context, allowing us to compare whether large-scale single-cell pretraining emphasizes the same metrics as LINCS pretraining.
The Tahoe training input follows the same OOD holdout policy: source compound identifiers remain available for audit, but molecule-level identities control training identity and leakage checks.
For both external resources, holdout filtering is performed against standardized molecule-level identities rather than source-specific compound labels.

\subsection{Baselines and Model Variants}

We compare against reference rows for CPA, chemCPA, and LINCS-pretrained chemCPA under the raw chemCPA preprocessing contract.
We also run the chemCPA-style backbone on our processed SciPlex3 benchmark without cell-cycle supervision.
These rows are denoted as chemCPA, no cell-cycle loss and serve as matched no-supervision baselines for the scCycleMol variants.
For scCycleMol, we evaluate variants that isolate the main design choices:
\begin{itemize}
    \item \textbf{Readout-only circular phase supervision}: the model uses the learnable cell-cycle head and circular phase loss on a stop-gradient copy of predicted treated expression.
    \item \textbf{Closed-loop circular supervision}: circular phase loss is allowed to update the upstream perturbation pathway.
\end{itemize}
We also evaluate embedding-source variants in the ablation section, comparing chemCPA embeddings with our molecular embeddings under LINCS and SciPlex pretraining.

\subsection{Evaluation Metrics}

Expression fidelity is measured with mean-expression $r^2$ over all modeled genes and over differentially expressed genes, averaged across perturbation conditions.
We also report the corresponding median $r^2$ values across conditions.
The DE-gene metric is important because it focuses evaluation on genes most responsive to the perturbation.
Phase accuracy is measured by phase classification accuracy from the predicted circular phase representation.
Models trained without $\losscc$ receive no phase-supervision gradient, and their predicted treated expression is scored post hoc by the same circular readout rule.
Their low phase accuracy reflects a chemCPA backbone evaluated without an explicit cell-cycle objective, so we interpret the phase score as a post hoc diagnostic.
Unless otherwise stated, the main tables report OOD results because this partition tests generalization to held-out molecules absent from SciPlex3 train/test and from the external pretraining inputs.

\subsection{Main Results}

\begin{table*}[t]
\centering
\small
\resizebox{\textwidth}{!}{%
\begin{tabular}{lllccccccc}
\toprule
Model & Pretraining & Preprocess & Mean $r^2$ all & Mean $r^2$ DEGs & $\Delta$ DEGs & Median $r^2$ all & Median $r^2$ DEGs & Phase acc. & $\Delta$ phase \\
\midrule
CPA & -- & Raw & 0.5000 & 0.2900 & -- & 0.4900 & 0.1200 & -- & -- \\
chemCPA & -- & Raw & 0.5100 & 0.3200 & -- & 0.4700 & 0.2400 & -- & -- \\
chemCPA & LINCS & Raw & 0.6800 & 0.5400 & -- & 0.7500 & 0.6400 & -- & -- \\
\midrule
chemCPA, no cell-cycle loss & -- & Ours & 0.8819 & 0.6315 & 0.0000 & 0.9294 & 0.7405 & 0.1918 & 0.0000 \\
scCycleMol, closed-loop circular & -- & Ours & 0.8806 & 0.6318 & \pos{+0.0003} & 0.9308 & 0.7449 & 0.7349 & \textbf{\pos{+0.5431}} \\
scCycleMol, readout-only circular & -- & Ours & 0.8829 & 0.6516 & \pos{+0.0201} & 0.9164 & 0.7080 & 0.8872 & \textbf{\pos{+0.6954}} \\
\midrule
chemCPA, no cell-cycle loss & LINCS & Ours & 0.9065 & 0.6695 & 0.0000 & 0.9369 & 0.7609 & 0.1928 & 0.0000 \\
scCycleMol, closed-loop circular & LINCS & Ours & 0.9085 & 0.6808 & \pos{+0.0113} & \textbf{0.9375} & \textbf{0.7774} & 0.8098 & \textbf{\pos{+0.6170}} \\
scCycleMol, readout-only circular & LINCS & Ours & \textbf{0.9093} & \textbf{0.6843} & \pos{+0.0148} & 0.9369 & 0.7595 & 0.9006 & \textbf{\pos{+0.7078}} \\
\midrule
chemCPA, no cell-cycle loss & Tahoe & Ours & 0.8801 & 0.6255 & 0.0000 & 0.9281 & 0.7293 & 0.1965 & 0.0000 \\
scCycleMol, closed-loop circular & Tahoe & Ours & 0.8824 & 0.6348 & \pos{+0.0093} & 0.9309 & 0.7439 & 0.8122 & \textbf{\pos{+0.6157}} \\
scCycleMol, readout-only circular & Tahoe & Ours & 0.8807 & 0.6312 & \pos{+0.0057} & 0.9315 & 0.7344 & \textbf{0.9609} & \textbf{\pos{+0.7644}} \\
\bottomrule
\end{tabular}
}
\caption{Out-of-distribution expression and cell-cycle results on SciPlex3. The first three rows are CPA/chemCPA reference baselines under the raw chemCPA preprocessing contract. The remaining rows use our processed molecule-aware SciPlex3 contract. $\Delta$ DEGs and $\Delta$ phase are computed relative to the matched chemCPA no-cell-cycle-loss row within the same pretraining group. Positive deltas are colored green. The no-cell-cycle rows are scored post hoc for phase accuracy, while closed-loop supervision updates the perturbation pathway and readout-only supervision trains a stop-gradient phase readout.}
\label{tab:main_results}
\end{table*}

\Cref{tab:main_results} separates preprocessing contracts from the cell-cycle objective.
The raw reference rows report CPA/chemCPA results under the original chemCPA preprocessing contract.
Under our processed molecule-aware contract, the matched chemCPA no-cell-cycle rows already achieve stronger expression metrics, with the LINCS-pretrained row reaching 0.9065 mean all-gene $r^2$ and 0.6695 mean DE-gene $r^2$.
The remaining comparisons isolate the added cell-cycle objective within the same processed-data setting.
Closed-loop circular supervision raises phase accuracy from about 0.19 to 0.73--0.81 while keeping mean all-gene $r^2$ within 0.003 of the corresponding chemCPA no-cell-cycle row.
Readout-only circular supervision reaches as high as 0.9609 phase accuracy, which indicates the strongest supervised phase readout from predicted expression.

\subsection{Dose-Resolved OOD Performance}

\begin{figure}[H]
\centering
\includegraphics[width=\linewidth]{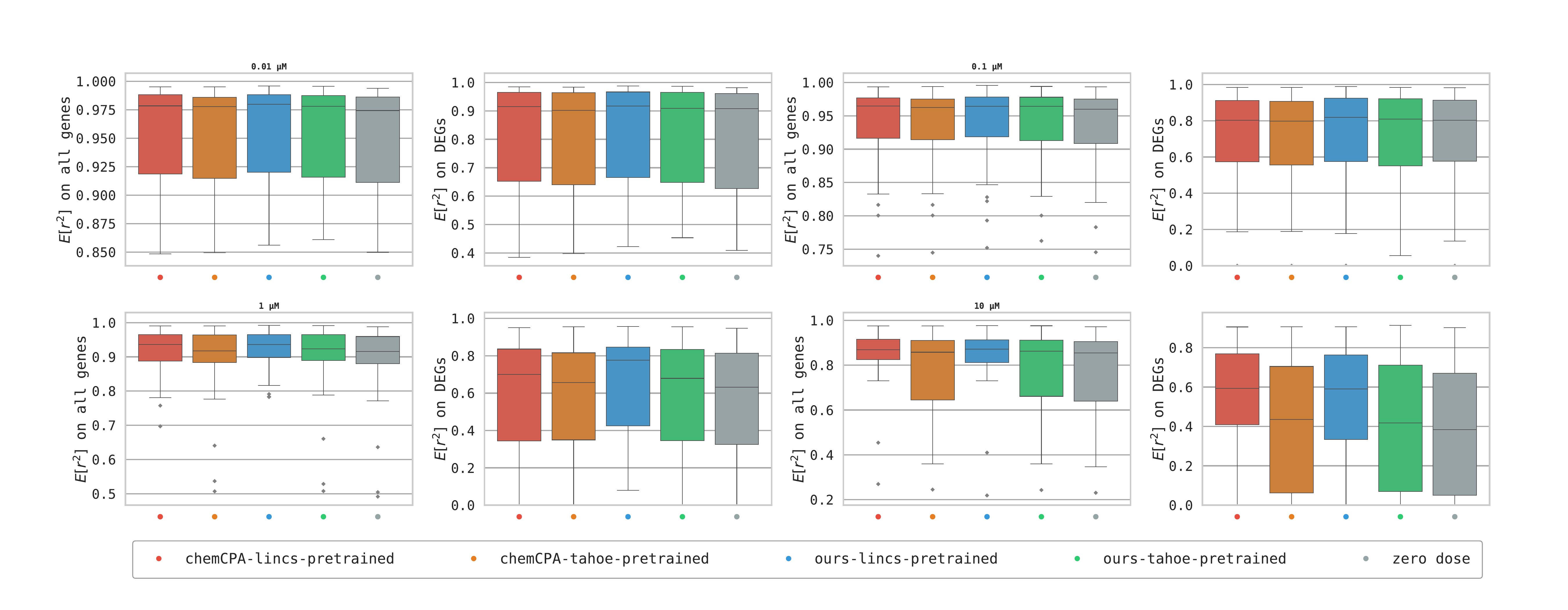}
\caption{Dose-resolved OOD $r^2$ distributions on SciPlex3. Each box summarizes cell-line--molecule combinations at one treatment dose for all modeled genes or DE genes. The zero-dose baseline predicts treated expression from matched DMSO controls.}
\label{fig:per_dose_boxplot}
\end{figure}

\Cref{fig:per_dose_boxplot} complements the aggregate OOD results by showing how performance varies across doses.
Across the all-gene panels, pretrained models and the zero-dose control have high median $r^2$ at low doses, indicating that much of the transcriptome remains close to the control state when perturbation effects are weak.
The DE-gene panels are more discriminative: their wider boxes and lower tails expose cell-line--molecule combinations where drug-responsive genes are substantially harder to predict.
Within matched pretraining settings, scCycleMol remains competitive with the processed-data chemCPA no-cell-cycle rows across doses, and the LINCS-pretrained scCycleMol variant has especially strong DE-gene medians in the mid-dose panels.
The dose-resolved view therefore supports the conclusion that the processed-data OOD expression performance is high on average and stable across the dose range where perturbation signal strength changes.

\subsection{Pretraining Effects}

\begin{figure}[H]
\centering
\includegraphics[width=0.62\linewidth]{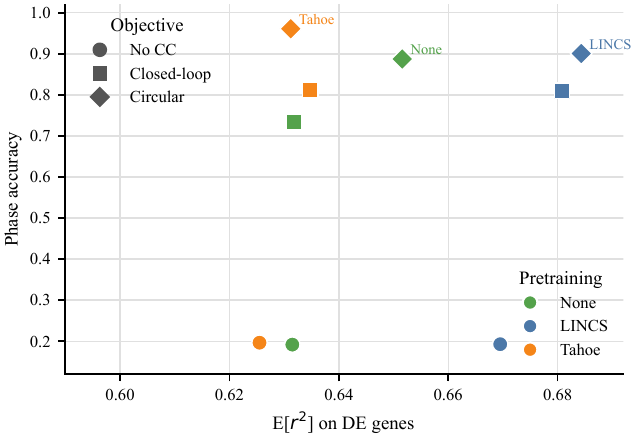}
\caption{Expression--cell-cycle trade-off across objectives and pretraining sources. Points farther right have better DE-gene expression prediction, while points higher have higher phase accuracy under the circular readout. LINCS pretraining moves models toward stronger expression metrics, while Tahoe pretraining gives the highest phase accuracy.}
\label{fig:tradeoff_scatter}
\end{figure}

As shown in \Cref{fig:tradeoff_scatter}, pretraining source changes the expression--cell-cycle trade-off.
LINCS pretraining gives the highest expression metrics, especially on DE genes, while Tahoe pretraining gives the highest phase accuracy.
We therefore treat pretraining source as an experimental factor rather than a universally dominant choice.

\subsection{Implementation Details}

All model variants use the same chemCPA-compatible train/test/OOD partitioning policy and the same evaluation protocol.
For fair comparison, the cell-cycle objective weight $\lambda_{\mathrm{cc}}$ is fixed within each ablation group before OOD evaluation.
The readout-only circular and closed-loop circular variants isolate whether the cell-cycle loss is used only to train a phase readout or is also allowed to update the upstream perturbation pathway.

\section{Analysis and Ablations}
\label{sec:analysis}

The main results separate expression prediction, closed-loop phase consistency, and supervised phase readout.
We next isolate which design choices support these outcomes under the processed-data contract.

\subsection{Closed-Loop Cell-Cycle Supervision}

Closed-loop supervision tests whether cell-cycle loss provides useful feedback to the perturbation pathway beyond a supervised phase readout.
We compare each matched chemCPA no-cell-cycle model with its closed-loop counterpart, where the circular phase loss backpropagates through the predicted treated expression into the upstream decoder and drug-conditioned perturbation pathway.

\begin{figure*}[t]
\centering
\begin{subfigure}[t]{0.48\textwidth}
    \centering
    \includegraphics[width=\linewidth]{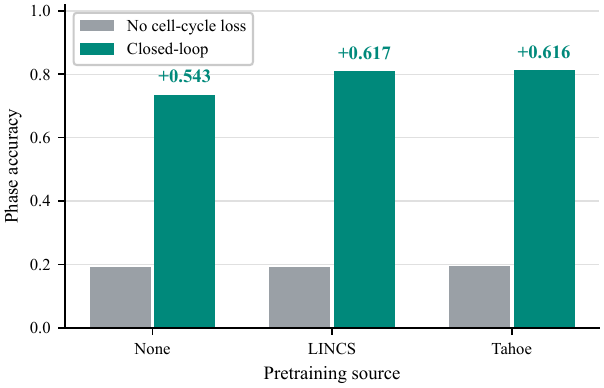}
    \caption{Closed-loop supervision.}
    \label{fig:closed_loop_gain}
\end{subfigure}
\hfill
\begin{subfigure}[t]{0.48\textwidth}
    \centering
    \includegraphics[width=\linewidth]{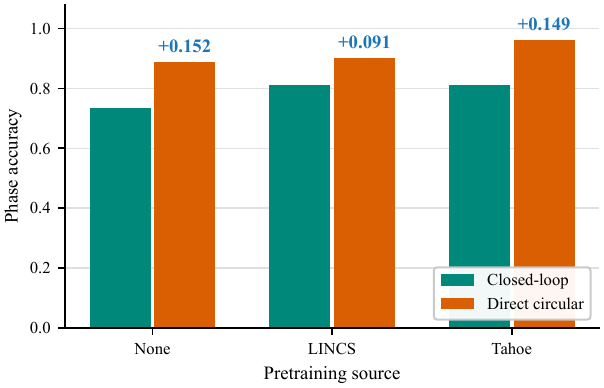}
    \caption{Closed-loop versus readout-only circular phase supervision.}
    \label{fig:closed_loop_vs_direct}
\end{subfigure}
\caption{Cell-cycle supervision ablations on the OOD split. (a) Allowing the circular phase loss to update the perturbation pathway improves phase accuracy by 0.54--0.62 absolute points across pretraining settings. (b) Readout-only circular phase supervision reaches the highest supervised phase readout accuracy, especially with LINCS or Tahoe pretraining.}
\label{fig:cc_ablation}
\end{figure*}

\Cref{fig:closed_loop_gain} supports the closed-loop design.
Across all three pretraining settings, phase accuracy increases sharply: from 0.1918 to 0.7349 without pretraining, from 0.1928 to 0.8098 with LINCS pretraining, and from 0.1965 to 0.8122 with Tahoe pretraining.
The corresponding changes in mean all-gene $r^2$ are small, ranging from -0.0013 to +0.0023.
Thus, the biological-state feedback improves phase consistency under the circular readout without introducing a large reconstruction penalty.

\subsection{Closed-Loop and Readout-Only Circular Phase Supervision}

Both circular variants use the learnable full-expression cell-cycle head.
They differ in whether the auxiliary signal is used only to train the phase readout or is also allowed to shape the perturbation pathway.
\Cref{fig:closed_loop_vs_direct} compares these alternatives.

Readout-only circular phase supervision gives the highest phase readout accuracy in all three pretraining settings.
It increases phase accuracy over the closed-loop variant from 0.7349 to 0.8872 without external pretraining, from 0.8098 to 0.9006 with LINCS pretraining, and from 0.8122 to 0.9609 with Tahoe pretraining.
This shows that the learnable full-expression head can extract phase information from predicted expression, while the feedback path and objective weight control the expression-versus-phase trade-off.

\subsection{Embedding Source Ablation}

We also examine whether the learned drug embedding improves OOD perturbation prediction beyond the corresponding chemCPA embedding.
\Cref{tab:embedding_ablation} compares chemCPA and scCycleMol embeddings under LINCS and SciPlex pretraining.

\begin{table}[H]
\centering
\small
\begin{tabular}{lcccc}
\toprule
Embedding setting & Mean $r^2$ all & Mean $r^2$ DEGs & Var. $r^2$ all & Var. $r^2$ DEGs \\
\midrule
LINCS + chemCPA embedding & 0.8756 & 0.5869 & 0.0361 & 0.0201 \\
LINCS + Ours embedding & \textbf{0.8841} & 0.6245 & 0.0510 & 0.0220 \\
SciPlex + chemCPA embedding & 0.8375 & \textbf{0.8011} & 0.7249 & 0.7230 \\
SciPlex + Ours embedding & 0.8386 & \textbf{0.8011} & \textbf{0.7257} & \textbf{0.7255} \\
\bottomrule
\end{tabular}
\caption{Embedding-source ablation on the OOD split. Under LINCS pretraining, our embedding improves mean all-gene and DE-gene $r^2$ over the chemCPA embedding. Under SciPlex pretraining, the two embeddings are nearly tied on mean DE-gene $r^2$, while our embedding slightly improves variance preservation.}
\label{tab:embedding_ablation}
\end{table}

The embedding ablation shows that representation gains depend on the pretraining source.
With LINCS pretraining, our embedding improves mean all-gene $r^2$ by 0.0085 and mean DE-gene $r^2$ by 0.0376 over the chemCPA embedding, with modest gains in both variance metrics.
With SciPlex pretraining, the two embedding choices are effectively tied on DE-gene response, while our embedding slightly improves all-gene mean $r^2$ and both variance metrics.
These results support using the proposed molecular embedding, but they also indicate that embedding choice should be interpreted together with the pretraining regime and the expression-versus-variance trade-off.

\subsection{Expression--Cell-Cycle Trade-Off}

The combined results suggest that expression reconstruction and phase accuracy should be reported together.
LINCS pretraining gives the strongest expression reconstruction, while Tahoe pretraining gives the highest phase accuracy.
Closed-loop supervision improves phase accuracy with little change in expression $r^2$, whereas readout-only circular supervision can maximize the supervised phase readout.
This trade-off motivates reporting both DE-gene $r^2$ and phase accuracy as primary metrics rather than optimizing only one of them.

\section{Conclusion}

We presented scCycleMol, a cell-cycle-aware framework for single-cell drug perturbation prediction.
The central idea is to use treated-state phase supervision as a biological-state signal: scCycleMol trains predicted treated expression to preserve circular G1/S/G2M phase targets through a learnable full-expression cell-cycle head.
This framing is supported by a processed cell-cycle-aware SciPlex3 benchmark and by molecular representation and pretraining ablations within the same prediction task.

Experiments on the processed SciPlex3 benchmark show strong OOD expression metrics under the molecule-aware preprocessing contract and separate these effects from the added cell-cycle objective.
Closed-loop circular supervision raises phase accuracy by 0.54--0.62 absolute points with little change in expression $r^2$, while readout-only circular phase supervision reaches the strongest supervised phase readout under Tahoe pretraining.
The embedding ablations show that drug representation gains depend on pretraining source, reinforcing the need to report expression response, variance preservation, and phase accuracy together.
Limitations include reliance on reliable phase labels, the coarse three-phase circular encoding, and the need to validate whether cell-cycle-aware supervision generalizes across broader perturbation datasets.
Future work should extend the framework to continuous cell-cycle trajectories and broader cellular state variables beyond cell cycle.

\bibliographystyle{plainnat}
\bibliography{references}

\appendix
\section{Additional Experimental Details}

\subsection{Preprocessing Overview}

We provide a reproducible Stage 1--2 preprocessing package for the three data resources used in this work: SciPlex3~\citep{srivatsan2020sciplex}, LINCS L1000~\citep{subramanian2017lincs}, and Tahoe-100M~\citep{zhang2025tahoe100m}.
Stage 1 prepares dataset-specific expression matrices, metadata, feature vocabularies, and quality-control outputs.
Stage 2 converts these outputs into model-ready training data, including compact expression matrices, cell and compound metadata, ChemBERTa drug embeddings, molecule-level perturbation identities, OOD holdout filtering, Top50 DEG annotations for SciPlex3, and validation manifests.
The migrated preprocessing code preserves the legacy transformation logic where appropriate, while updating the formal data contract to use standardized-molecule identities rather than source-specific compound labels.
Legacy-compatible outputs may be retained for auditability, but the formal paper protocol uses the molecule-aware, Top50-DEG, OOD-holdout-filtered assets described below.
For model-facing identities, each dataset uses the standardized molecule SMILES emitted by its preprocessing path; external L1000 and Tahoe assets additionally use the shared molecule-identity helper, where the model SMILES is RDKit canonical isomeric SMILES.
For cross-resource leakage control, the held-out SciPlex3 OOD molecules are matched more broadly using audit keys that include canonical and non-stereo SMILES, fragment/charge/tautomer parent SMILES, and InChIKey or connectivity keys.
This broader audit matching is used only to remove potentially overlapping compounds from external pretraining inputs; source compound identifiers remain traceability fields.

The public release boundary is the preprocessing code, configuration files, lightweight documentation, and static manifest summaries.
Raw datasets, third-party pretrained model weights, large processed outputs, local run directories, and per-run manifest JSON files are not redistributed.
Users are expected to download raw datasets from the official sources and regenerate processed outputs locally.
ChemBERTa embedding steps use a local copy of \texttt{seyonec/ChemBERTa-zinc-base-v1}; model weights are not included in the repository.

\subsection{Dataset Statistics After Preprocessing}

\begin{table}[H]
\centering
\small
\resizebox{\textwidth}{!}{%
\begin{tabular}{lrrrrl}
\toprule
Dataset & Cells / samples & Model perturbation identities & Genes & Cell lines & Formal asset \\
\midrule
SciPlex3 & 635,541 cells & 186 non-control molecule units & 5,080 & 3 & molecule-aware Top50-DEG AnnData \\
Tahoe-100M & 92,670,735 cells & 351 non-control molecule units & 5,037 & 49 & OOD-holdout-filtered training shards \\
LINCS L1000 & 157,927 signatures & 20,388 molecule IDs & 978 & 21 & OOD-holdout-filtered Stage-0 dataset \\
\bottomrule
\end{tabular}
}
\caption{Formal model-ready assets generated by the Stage 1--2 preprocessing pipelines. The perturbation-identity column reports model-facing molecule-level units rather than source compound labels; source identifiers are retained separately for traceability. SciPlex3 has 186 non-control molecule units, and Tahoe has 351 non-control molecule units, or 352 molecule identities when the control identity is included. The external L1000 and Tahoe assets are filtered against the SciPlex3 OOD molecule list using molecule audit keys.}
\label{tab:preprocess_dataset_stats}
\end{table}

\subsection{SciPlex3 Processing}

SciPlex3 processing starts from the decompressed GEO large-screen files for A549, MCF7, and K562.
The required Stage 1 path performs basic quality control, hash-based doublet removal, highly variable gene selection, and cell-cycle scoring.
The required Stage 2 path extracts training features, computes ChemBERTa embeddings for standardized SMILES strings, writes the compact training AnnData object, assigns molecule-aware splits, and annotates Top50 perturbation-responsive genes for formal DE-gene evaluation.

\begin{table}[H]
\centering
\small
\resizebox{\textwidth}{!}{%
\begin{tabular}{p{0.32\linewidth}rrp{0.28\linewidth}}
\toprule
Step & Input & Output & Main operation \\
\midrule
Basic filtering & 799,317 cells, 58,347 genes & 661,596 cells, 57,591 genes & Remove incomplete metadata, non-target time points, low UMI/gene cells, high mitochondrial fraction, and cleaned-up genes. \\
Doublet filtering & 661,596 cells & 635,541 cells & Remove 26,055 hash doublets. \\
HVG selection & 57,591 genes & 5,080 genes & Select HVGs with legacy-equivalent settings and force-include cell-cycle genes. \\
Cell-cycle scoring & 635,541 cells & 635,541 cells & Add S score, G2/M score, phase label, phase ID, and proliferation score. \\
Feature extraction & 635,541 cells & 186 molecule-level canonical SMILES & Add standardized SMILES, dose, cell-line, time, and control indicators. \\
ChemBERTa embedding & 189 compound labels & 188 valid embeddings + 1 control sentinel & Encode valid label-level compounds with 768-dimensional ChemBERTa CLS embeddings; write a zero-vector sentinel for the control/invalid placeholder and attach the resulting per-cell matrix. \\
Training object & 635,541 cells, 5,080 genes & 635,541 cells, 5,080 genes & Write compact AnnData with 9 obs columns, \texttt{counts}, \texttt{log1p\_norm}, \texttt{compound\_chemberta}, and unified gene token IDs. \\
Molecule-aware split assignment & 635,541 cells & 564,501/45,862/25,178 train/test/OOD cells & Add the chemCPA-compatible OOD split over 186 non-control molecule units and an optional molecule-strict split for sensitivity analysis. \\
Top50 DEG annotation & 635,541 cells, 5,080 genes & 2,232 non-control groups & Compute the Top50 perturbation-responsive genes for each non-control cell-line--molecule--dose group. \\
\bottomrule
\end{tabular}
}
\caption{SciPlex3 preprocessing path under the molecule-aware Top50-DEG contract. The model-facing perturbation identity is the molecule-level condition; label-level embedding entries are retained only as molecular representation inputs and traceability links. Cell-cycle scores matched the legacy output within $2.9\times 10^{-7}$ maximum absolute difference, and phase labels matched exactly.}
\label{tab:sciplex_preprocessing}
\end{table}

\paragraph{Molecule identity and repeated structures.}
Standardized SMILES define the molecule-level perturbation identities used for training and split assignment.
Original SciPlex3 product labels, LINCS perturbation identifiers, and Tahoe compound identifiers are retained as traceability metadata rather than used as canonical drug identities.
Repeated standardized structures are not averaged or dropped: sample rows remain distinct observations across cell line, dose, batch, and replicate, while shared structures map to the same molecule-level identity.
For external leakage checks, the molecule-level identity is supplemented with broader parent, tautomer, stereochemistry-insensitive, and InChIKey-based audit keys before filtering L1000 or Tahoe.

\paragraph{Top50 DEG contract.}
SciPlex3 DE-gene evaluation uses the top 50 perturbation-responsive genes for each non-control cell-line--molecule--dose group, producing 2,232 groups with 50 genes each.
Control reference groups are excluded from the DEG dictionary.
Legacy full-gene DEG annotations are retained only for compatibility and are not used for paper-style DE-gene evaluation.

\subsection{Tahoe-100M Processing}

Tahoe-100M processing starts from public expression parquet shards and metadata tables.
The pipeline validates shard continuity and metadata, computes gene statistics, selects HVGs, extracts sample/compound/cell-line/dose/phase features, creates training-ready expression shards, embeds compounds with ChemBERTa, and validates the resulting package.
Normal cell lines are excluded from tumor-only statistics and training data.
Compounds without a compatible single-compound SMILES representation are removed from the feature and training outputs.
The formal Tahoe asset additionally removes cells whose compound rows match the SciPlex3 OOD molecule audit-key set; source compound identifiers remain available for audit, but molecule-level conditions control training identity and split assignment.
The Tahoe HVG step uses a scalable Seurat-v3-inspired mean/variance procedure with forced cell-cycle genes rather than a full official Scanpy \texttt{seurat\_v3} run on all cells.

\begin{table}[H]
\centering
\small
\resizebox{\textwidth}{!}{%
\begin{tabular}{p{0.32\linewidth}rrp{0.28\linewidth}}
\toprule
Step & Input & Output & Main operation \\
\midrule
Data validation & 3,388 shards; 100,648,790 metadata rows & Validation reports & Check parquet continuity, metadata readability, barcode coverage, and sampled expression rows. \\
Gene statistics & 62,710 genes & 54,877 detected genes & Compute gene-level detection and expression statistics after excluding normal-cell summaries. \\
HVG selection & 62,710 genes & 5,037 HVGs & Use a scalable Seurat-v3-inspired HVG procedure, remove 40,337 genes by HVG prefilter, and force-add 37 cell-cycle genes. \\
Input features & 1,344 samples & 1,341 samples; 379 source compounds; 354 molecule identities; 49 cell lines & Remove three samples from an incompatible multi-compound drug representation and build feature vocabularies. \\
Training data & 95,624,334 cells & 93,698,301 cells & Remove normal-cell-line cells after prior incompatible-drug/sample filtering; write HVG shards and cell metadata. \\
OOD holdout filtering & 93,698,301 cells; 379 source compounds & 92,670,735 cells; 376 source compounds; 351 non-control molecule units & Remove cells from compound rows matching the SciPlex3 OOD molecule audit-key set. \\
ChemBERTa embedding & 379 source compounds & 376 embedding entries (375 filtered + 1 control sentinel) & Encode compatible source compounds before holdout filtering; copy the filtered embedding sidecar, retaining one zero-vector sentinel in the formal asset. \\
Validation & training-ready outputs & 50/50 passed checks & Validate required fields, vocabularies, metadata, embeddings, expression shards, excluded-compound removal, and training manifest. \\
\bottomrule
\end{tabular}
}
\caption{Tahoe-100M preprocessing path. The formal OOD-holdout-filtered asset contains 92,670,735 cell metadata rows after removing 1,027,566 cells from three Tahoe compound IDs matched to SciPlex3 OOD molecules. After source-compound filtering, 376 Tahoe compound rows map to 351 non-control molecule-level split units, or 352 molecule identities when the control identity is included; molecule split counts are 281 train, 35 test, and 35 OOD non-control units.}
\label{tab:tahoe_preprocessing}
\end{table}

\subsection{LINCS L1000 Processing}

L1000 processing creates a Stage 0 VAE dataset from treatment and control signatures.
The pipeline filters signatures by perturbation type, tumor sample type, treatment duration, usable canonical SMILES, quality metrics, and required Stage 0 fields.
It then extracts landmark-gene expression matrices for treatment and control signatures, computes ChemBERTa embeddings for compounds, and joins treatment signatures with matched expression, compound embeddings, and cell-line vehicle controls.
The formal Stage 0 asset is filtered against the SciPlex3 OOD molecule audit-key set before pretraining.
After filtering, it retains 20,714 source perturbation identifiers corresponding to 20,388 molecule-level identities.
LINCS source perturbation identifiers are retained for traceability, while molecule-level identifiers define condition identity, split units, and repeated-structure weighting.

\begin{table}[H]
\centering
\small
\resizebox{\textwidth}{!}{%
\begin{tabular}{p{0.34\linewidth}rrp{0.26\linewidth}}
\toprule
Step & Input & Output & Main operation \\
\midrule
Load and filter & 591,697 signatures & 158,151 treatment signatures; 8,444 controls & Remove non-\texttt{trt\_cp}, non-tumor, incompatible duration, unusable SMILES, low-quality, or incomplete signatures. \\
Treatment expression & 158,151 signatures & $158{,}151 \times 978$ matrix & Extract L1000 landmark-gene expression for treatment signatures. \\
Control expression & 8,444 signatures & $8{,}444 \times 978$ matrix & Extract matched control expression for vehicle signatures. \\
ChemBERTa embedding & 20,721 source perturbation IDs & $20{,}721 \times 768$ matrix & Standardize SMILES and compute source-level compound embeddings before OOD holdout filtering. \\
Stage 0 dataset & 158,151 treatment signatures & 158,151 samples & Join expression, compound embeddings, metadata, control means, genes, condition summaries, and split units. \\
OOD holdout filtering & 158,151 samples & 157,927 samples; 20,388 molecule IDs & Remove 224 signatures from seven source perturbation identifiers matching the SciPlex3 OOD molecule audit-key set; split molecule IDs into 16,310 train, 2,039 test, and 2,039 OOD units. \\
\bottomrule
\end{tabular}
}
\caption{LINCS L1000 preprocessing path. The formal OOD-holdout-filtered Stage-0 asset contains 157,927 samples and 20,388 molecule-level identities; 20,714 source perturbation identifiers are retained only for traceability.}
\label{tab:l1000_preprocessing}
\end{table}

\subsection{Gene Vocabulary and Manifests}

All datasets use a shared lightweight gene vocabulary for stable token joins.
The public key table contains stable Ensembl-to-token mappings, final-dataset membership flags, retired-ID replacement fields, and traceability annotations.
Only \texttt{Unified\_ENSG} and \texttt{Unified\_Token\_ID} are stable join keys; gene symbols are annotations and may reflect dataset-original, alias, or retired-ID names.
Rows marked as traceability-only are retained for auditability and are not additional training genes.

Each accepted preprocessing step writes a machine-readable manifest recording inputs, outputs, filtering decisions, counts, configuration values, runtime, and hashes where practical.
The public documentation includes a static cross-dataset manifest summary generated from the accepted manifests.
The manuscript reports the corresponding data contracts rather than local run commands or site-specific paths.

\end{document}